\documentstyle[12pt,twoside,fleqn,espcrc1]{article}

\newcommand{\AmS}{{\protect\the\textfont2
  A\kern-.1667em\lower.5ex\hbox{M}\kern-.125emS}}

\title{
Beam Energy Evolution of HBT Systematics at the AGS
}

\author{M.A. Lisa$^+$, for the E895 Collaboration\thanks{
Author list and Grant acknowledgements may be found in G. Rai's contribution to these Proceedings.}
\\
$^+$Physics Department, Ohio State University, 174 W. 18th Ave., Columbus, OH 43210}

\begin{document}
\maketitle

\begin{abstract}
We present preliminary results of the first $\pi$
interferometry (HBT) excitation function at intermediate AGS energies. 
The beam energy evolution of the correlations' dependence on $m_T$, centrality, and
emission angle with respect to the reaction plane are discussed.  Comparisons
with predictions of the RQMD cascade model are made.
\end{abstract}

\vspace*{0.2in}

Two-particle intensity interferometry (HBT) measurements have long been used to
study the geometry and dynamics of heavy ion collisions (see,
e.g. \cite{bauer_gelbke_pratt}).  Pion correlation functions
are sensitive to the pion source size, shape, decay-time, and
long-lived particle (e.g. $\Lambda$) production.  In addition,
dynamic effects such as flow produce space-momentum correlations resulting in
dependences of the correlation functions on $\pi$ momentum.

In this paper, we discuss  an excitation function (2-8
AGeV) of $\pi^-$ HBT measurements.  Studying the evolution of the correlations
as a function of $E_{beam}$ is important for two reasons.
Firstly, a sudden increase, at some $E_{beam}$, in the lifetime
of the hadronic fireball has long been proposed as a robust signal of the onset
of QGP formation~\cite{pratttimescale,bertsch-pratt,RischkeHBT}.
Secondly, the sensitivity of correlation functions to the underlying physics
makes such measurements potent tools to test the dynamics of
microscopic models of heavy ion collisions.
Many models attempt to extrapolate to the RHIC energies.
Confidence in the ability to extrapolate
(determined by the correct underlying physics and its evolution with energy)
would be enhanced if the model
reproduces an {\em excitation function} of detailed HBT systematics.

Using the large-acceptance EOS Time Projection Chamber~\cite{EOSTPC}
the E895 collaboration measured charged particles from Au+Au collisions at
2, 4, 6, and 8 AGeV at the Brookhaven AGS.
Good particle identification
minimized $e^-$ contamination of the $\pi^-$ sample.  Momentum resolution,
largely due to multiple Coulomb scattering and straggling in the 3\% interaction
length target, was on the order of 1.5-3\%.  The experimental correlation functions 
have been corrected for the momentum resolution with an iterative method similar
to that employed by the NA44 collaboration~\cite{NA44-accept}.  This correction
typically increases the fitted $\lambda$ parameter by 15\%, and the radii by 5\%.

Track merging and splitting effects were eliminated
by a two-track geometrical cut based on the tracking algorithm.
As expected, this cut discards some pairs (in the ``real'' and event-mixed
distributions) at low relative momentum, q.
However, due to detector geometry, this cut preferentially discards
low-q pairs at high $p_T$; thus, to minimize phasespace bias effects, we
restrict our analysis to low $p_T$ and use narrow windows in $p_T$.

A full Coulomb-wave integration~\cite{pratt_csorgo,Messiah}
over a spatial source of 5-fm Gaussian radius
was used to generate the Coulomb correction, which was applied pair-wise to
the event-mixed denominator of the correlation function.  Identical Coulomb
correction functions were applied to data and to correlation functions from the
RQMD (see below).
Further details of the analysis have been reported previously~\cite{lisa_parkcity99}.

\begin{figure}[bh]
\vspace*{2.0in}
\includegraphics{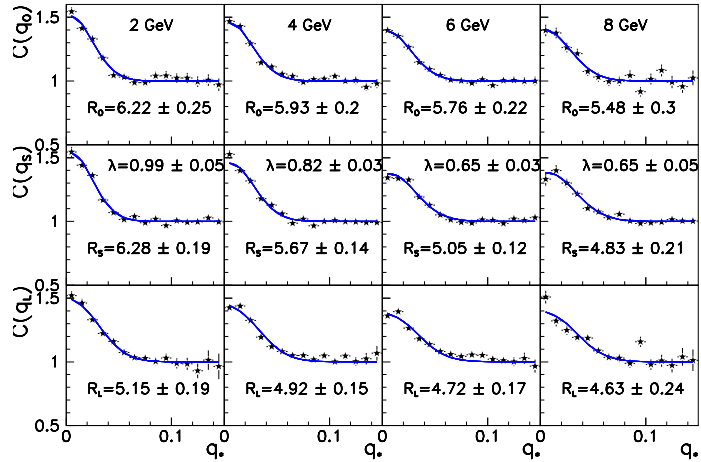}
\caption{
Projections of the 3-D $\pi^-$ correlation functions from central collisions ($\sim$11\%$\sigma_T$).
$q_{out}$ (top), $q_{side}$ (middle), and $q_{long}$ (bottom) projections are shown
for collisions at all beam energies measured.
$y=y_{cm}\pm$0.35 and $p_T$=0.1-0.3 GeV/c cuts were applied.  Projections
in a given q-component are integrated over $\pm$30~MeV/c in the other components.
Superimposed curves show projections of 3-D fits described in text.
}
\label{fig:projections}
\vspace*{-0.25in}

\end{figure}

The high quality of the data is seen in Figure~\ref{fig:projections}, where projections
in the Bertsch-Pratt (BP), or out-side-long, system are shown for midrapidity pions from
central events at each bombarding energy.  The relative momentum $q$ was calculated in the fixed
Au+Au c.m. frame.  The functional form

\begin{equation}
\label{eq:bertsch-pratt}
C(q_{out},q_{side},q_{long})=1+
\lambda e^{-R_{out}^2q_{out}^2
-R_{side}^2q_{side}^2-R_{long}^2q_{long}^2-2R_{ol}^2q_{out}q_{long}}
\end{equation}

\noindent
was fit to the data, using a maximum-likelihood technique~\cite{E877QM96}.  The
cross-term~\cite{chapman-hienz} $R^2_{ol}$ was consistent with zero in all cases
 and uncorrelated with the other parameters.

Correlation functions were also constructed in the 
Yano-Koonin-Podgoretski\u{i} (YKP) decomposition~\cite{YKP}; 
here, the effective lifetime $R^0$ is fit more directly.  We fit to the form

\begin{equation}
\label{eq:yano-koonin}
C(q_{0},q_{\perp},q_{\parallel})=1+\lambda e^{-R_{\perp}^2q_{\perp}^2
-R_{\parallel}^2(q_{\parallel}^2-q_{0}^2)-
(R_{o}^2+R_{\parallel}^2)(q\cdot U)^2}
\end{equation}

\begin{figure}[t]
\vspace*{2.3in}
\begin{minipage}[t]{80mm}
\includegraphics{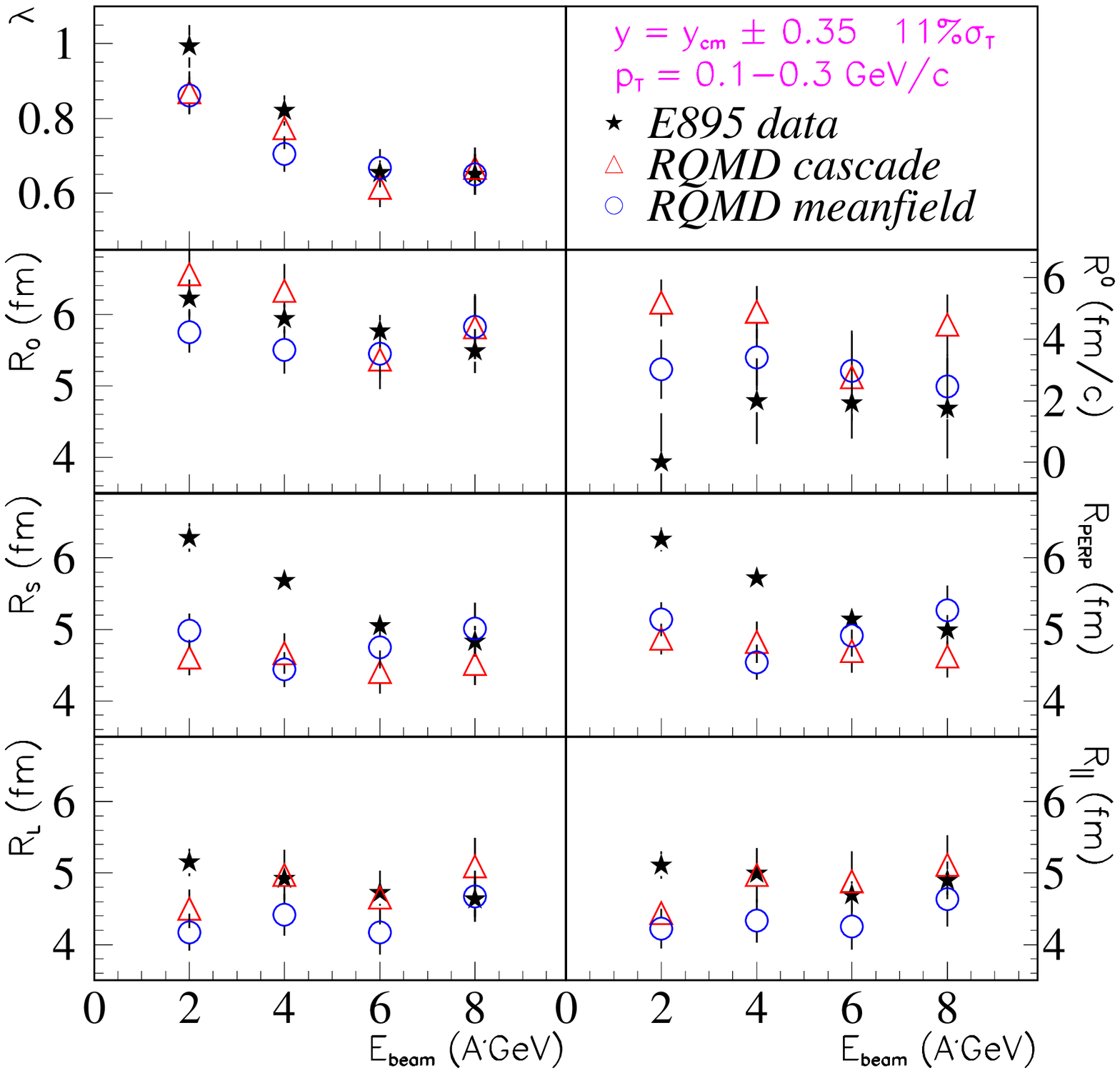}
\caption{
Beam energy dependence of the B-P and YKP fit parameters.
}
\label{fig:fit-excitation}
\end{minipage}
\hspace{\fill}
\begin{minipage}[t]{75mm}
\includegraphics{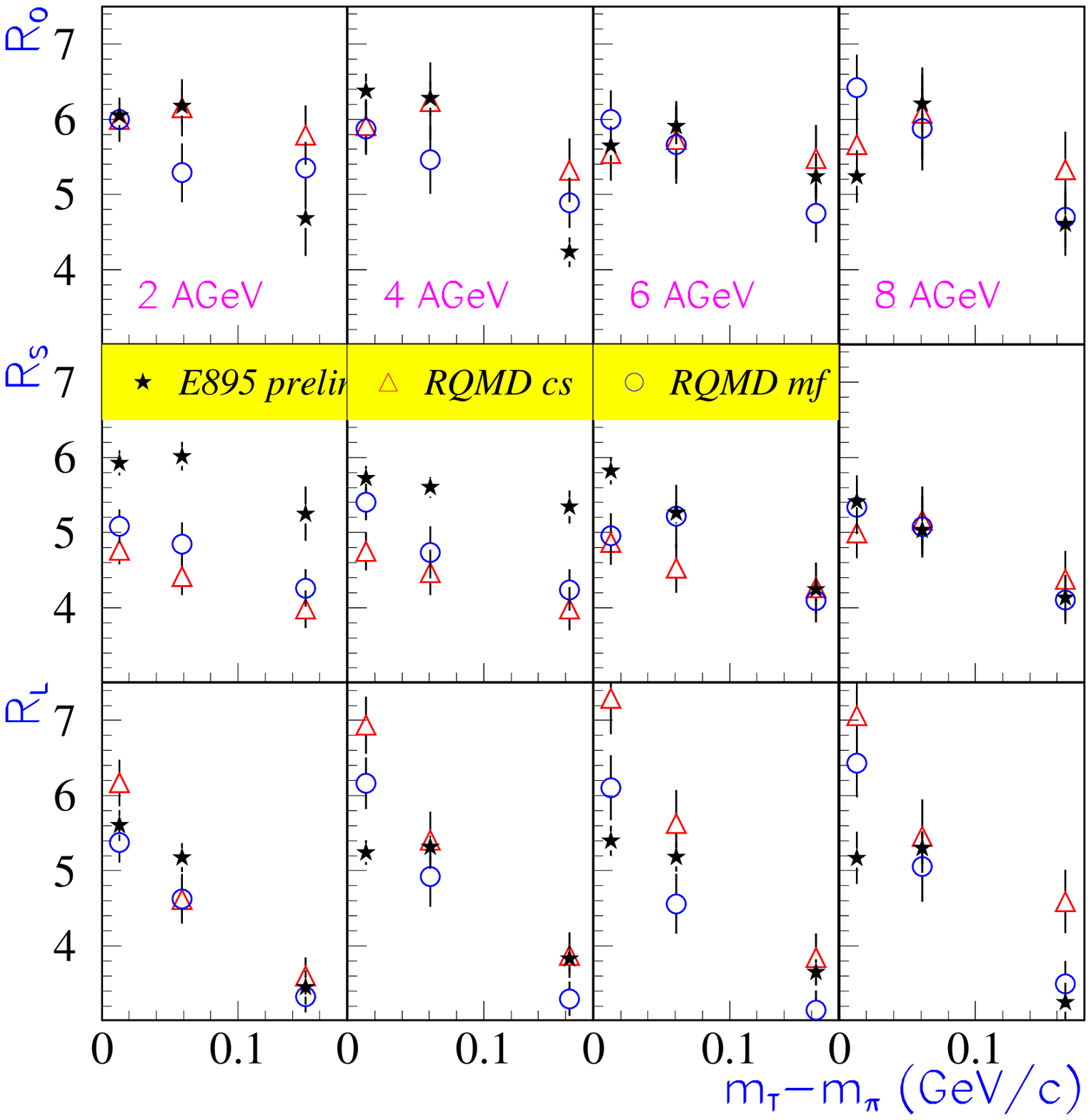}
\caption{
$m_T$ dependence of the B-P radii for data and RQMD.
}
\label{fig:mt-dep}
\end{minipage}
\vspace*{-0.15in}
\end{figure}

The excitation function of the fit results is presented in Figure~\ref{fig:fit-excitation}.
Also shown are results of fits to correlation functions generated by using the $\pi^-$ freeze-out points
from the RQMD (v2.3) model~\cite{rqmd} as input to the two-particle correlator code
CRAB~\cite{pratt-CRAB}.  Both the data and the model show a decrease in the $\lambda$ parameter,
due to increased production of long-lived $\pi^-$-emitting particles at higher
energy~\cite{pratt_csorgo}.

While the longitudinal radii, $R_{long}$ and $R_{\parallel}$, display little dependence on beam energy,
the observed decrease of the transverse radii $R_{side}$ and $R_{out}$ comes as something of a surprise.
At low energy, the YKP fits suggest that the model produces a $\pi^-$ source that is too small and too
long-lived; this leads in the BP decomposition to an underprediction in $R_{side}$, but
a reasonable agreement in $R_{out}$, as the space and time effects partially cancel.

Part of the reason for the decrease in the apparent transverse size is revealed in Figure~\ref{fig:mt-dep}.
It is clear that the $m_T$-dependence of $R_{side}$ (and the space-momentum correlation that causes it)
becomes stronger with $E_{beam}$ in the data, while
the model suggests a $m_T$-dependence rather independent of energy, and reminiscent of trends at higher
energy.  The observed trend may suggest that collective transverse flow
builds with bombarding energy, and is only strong enough to affect $R_{side}(m_T)$ above $E_{beam}\sim$4~AGeV.

\begin{figure}[htb]
\vspace*{2.3in}
\begin{minipage}[t]{80mm}
\includegraphics{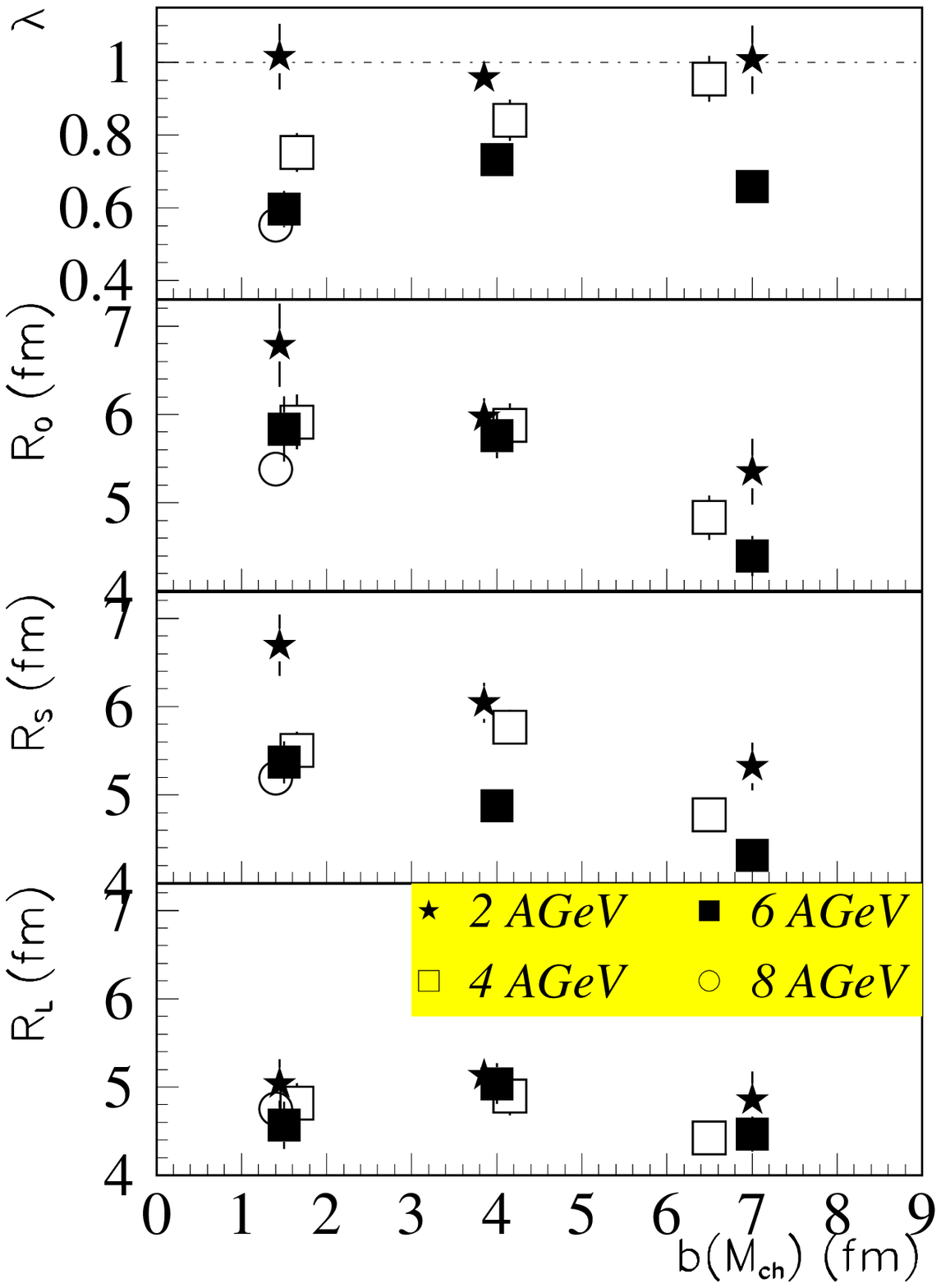}
\caption{
Centrality dependence of the BP fit parameters.
}
\label{fig:centrality}
\end{minipage}
\hspace{\fill}
\begin{minipage}[t]{75mm}
\includegraphics{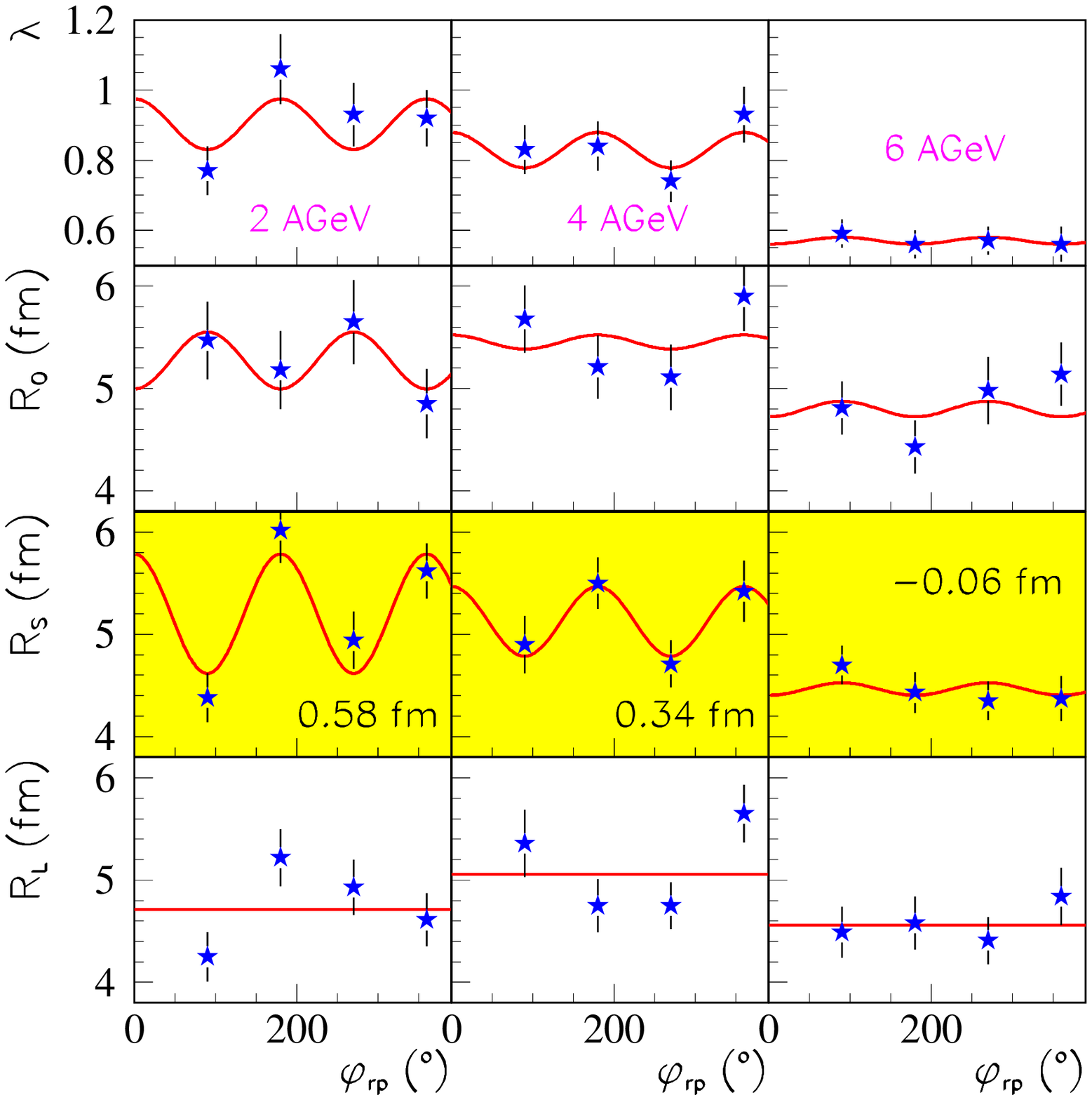}
\caption{
BP parameters versus $\pi^-$ emission angle with respect to reaction plane.
}
\label{fig:rp_dep}
\end{minipage}

\vspace*{-0.15in}
\end{figure}

Although dynamics determines the $m_T$-dependence of the radii, it is worthwhile to check that the
HBT radii track somewhat with geometry.  Figure~\ref{fig:centrality} shows the impact parameter ($b$)
dependence of the HBT radii ($b$ was estimated from the charged particle multiplicity).
$\lambda$ increases with $b$, as the production
of long-lived $\pi$-emitting particles is suppressed relative to direct pions.
The transverse radii ($R_{side}$ and $R_{out}$) decrease for more peripheral collisions, as expected, while
$R_{long}$ shows little $b$-dependence.
The trends suggest that the measured pion
source reflects the overlap volume of the colliding nuclei.

More detailed
information may be obtained by studying the HBT signal as a function
of $\pi^-$ emission angle with respect to the reaction plane.
The reaction plane is
calculated only from momenta of Z$\leq2$ nuclei,
for every event~\cite{E895-ellipticflow}, so auto-correlations are not an issue.
From an overlap-volume picture,
one expects an anisotropic
apparent shape in the transverse direction for non-central collisions, with a
larger spatial scale perpendicular to the reaction plane.
Deviations may reflect the non-isotropic
flow dynamics of the system prior to freeze-out, or may carry
information concerning the opacity of the source~\cite{heiselberg}.

Preliminary results for 2, 4, and 6 AGeV collisions at $b\approx5-7$ fm
are shown in Fig.~\ref{fig:rp_dep}.
At the lower energy, $R_{side}$ (the radius most closely related to geometry~\cite{pratttimescale})
exhibits a $\phi_{rp}$-dependence consistent with geometric considerations. While RQMD simulations 
(with perfect reaction plane resolution) display
similar trends at all energies,
at higher energy, the $R_{side}$ oscillation is not seen in our data.
Since the radii are not corrected for the finite dispersion, this is due at least in part to the worsening
resolution with which the reaction plane is measured.
Further study of this novel HBT signal is required.

In summary, we are mapping out the systematics of pion correlations in the energy range
between the Bevalac and maximum AGS energy.  Large jumps in source size or lifetime at some
collision energy, which might indicate the onset of QGP formation, are not observed.
Surprisingly, the apparent source size is larger at the lower beam energies; this appears
largely a consequence of weaker space-momentum correlations there.  The RQMD model, with or without
meanfield effects, does not reproduce the data; in the model at low energy, the effective size
is too small, the lifetime too large, and $R(m_T)$ does not
evolve with beam energy.  The impact parameter dependence
of the radii follows  naive expectations from geometry.  An HBT analysis correlated with
the event-wise reaction plane reveals a significant oscillation in $R_{side}$, at low beam energy.



\end{document}